\documentclass[preprint,showpacs,preprintnumbers,amsmath,amssymb]{revtex4}
\usepackage{graphicx}% Include figure files
\usepackage{dcolumn}% Align table columns on decimal point
\usepackage{bm}% bold math
\begin{document}

\title{Critical points in the Bragg glass phase of a weakly pinned crystal of Ca$_3$Rh$_4$Sn$_{13}$}
 
\author{S. SARKAR$^1$, A.D.  THAKUR$^1$, C.V. TOMY$^{2,\ast}$, G. BALAKRISHNAN$^3$, D.McK. PAUL$^3$, S. RAMAKRISHNAN$^1$ and A. K.  GROVER$^{1,\dagger}$}

\affiliation{$^1$ DCMP\&MS, Tata Institute of Fundamental Research, Mumbai
400 005, India  \\
$^2$ Department of Physics, Indian Institute of Technology Bombay, Mumbai
400 076, India  \\
$^3$ Department of Physics, University of Warwick, Coventry CV4 7AL, UK}
\date{today}

\begin{abstract}
New experimental data are presented on the scan rate dependence of the magnetization hysteresis width $\Delta M(H)$ ($\propto$ critical current density $J_c(H)$) in isothermal $M-H$ scans in a weakly pinned single crystal of Ca$_3$Rh$_4$Sn$_{13}$, which displays second magnetization peak (SMP) anomaly as distinct from the peak effect (PE). We observe an interesting modulation in the field dependence of a parameter which purports to measure the dynamical annealing of the disordered bundles of vortices injected through the sample edges towards the destined equilibrium vortex state at a given $H$. These data, in conjunction with the earlier observations made while studying the thermomagnetic history dependence in $J_c(H)$ in the tracing of the minor hysteresis loops, imply that the partially disordered state heals towards the more ordered state between the peak field of the SMP anomaly and the onset field of the PE. The vortex phase diagram in the given crystal of  Ca$_3$Rh$_4$Sn$_{13}$ has been updated in the context of the notion of the phase coexistence of the ordered and disordered regions between the onset field of the SMP anomaly and the spinodal line located just prior to the irreversibility line. A multi-critical  point and a critical point in the ($H,T$) region of the Bragg glass phase have been marked in this phase diagram and the observed behaviour is discussed in the light of recent data on multi-critical point in the vortex phase diagram in a single crystal of Nb.  

\end{abstract}

\keywords{Second Magnetization Peak, ramp rate dependence of hysteresis width,Peak effect, order-disorder transition, Ca$_3$Rh$_4$Sn$_{13}$, multi-critical point, critical point, vortex phase diagram.}

\pacs{74.25.Qt, 64.70.Dv, 74.25.Dw, 74.25.Sv}

\maketitle

\section{Introduction}

The vortex phase diagram in the Ca containing ternary stannide, Ca$_3$Rh$_4$Sn$_{13}$ compound attracted attention \cite{1} after it was reported \cite{2,3} that the peak effect (PE) phenomenon in the critical current density ($J_c$) in single crystals of isostructural Yb$_3$Rh$_4$Sn$_{13}$ compound bore strong resemblance to that in the U-based intermetallics, UPd$_2$Al$_3$ \cite{4} and UPt$_3$ \cite{5}. In the latter systems, a case had been made \cite{4} for the creation of the Fulde-Ferrel-Larkin-Ovchinnikov (FFLO) state  \cite{6,7} leading to the onset of the PE anomaly. Yb, being a rare earth element, has the potential of imparting adequately high paramagnetic susceptibility to Yb$_3$Rh$_4$Sn$_{13}$ even while it is in the mixed valent mode (Yb$^{3+}$ to Yb$^{2+}$). The Yb$_3$Rh$_4$Sn$_{13}$ compound could thus emerge as a possible candidate for the realization of the FFLO state. Initial studies by Sato {\it et al} \cite{2} and Tomy {\it et al} \cite{1,3} had indicated that the PE in the ternary stannides was observable at $H > 7$~kOe and for $T/T_c(0) < 0.8$. However, the latter studies \cite{8,9,10,11} in single crystals of Ca$_3$Rh$_4$Sn$_{13}$ and Yb$_3$Rh$_4$Sn$_{13}$ revealed the occurrence of PE in isofield ac susceptibility ($\chi^{\prime}(T)$) scans in fields down to about 3~kOe and reduced temperature $T/T_c(0)$ value upto 0.94. The $\chi^{\prime}(T)$ scans in these compounds also revealed \cite{8,10,11} the occurrence of notion of stepwise disordering of the ordered vortex state across the temperature interval extending from the onset position of the PE upto the sharp commencement of depinning process in the bulk of the sample. Prior to this work, such a behavior had been reported in weakly pinned crystals of the systems like, $2H$-NbSe$_2$ and the cubic ($C$15) CeRu$_2$ \cite{12}. The underlying intrinsic physics of the layered $2H$-NbSe$_2$, cubic ($C$15) CeRu$_2$ and the ternary stannides may be different in different systems. This in turn implies that a common basis of the detailed behavior across the PE region in different compounds presumably lies in the disordering of the elastically deformed ordered vortex state in response to the competition between the interaction effect between the vortices and the effective pinning, while the pinning and the elasticity of the vortex lattice rapidly start to collapse in a different manner, while approaching the superconducting-normal phase boundary \cite{13}.

A characteristic feature which distinguishes the vortex state of the single crystal sample of Ca$_3$Rh$_4$Sn$_{13}$ \cite{11} (in a portion of the ($H, T$) phase space) from those in the crystals of Yb$_3$Rh$_4$Sn$_{13}$ \cite{10}, CeRu$_2$ \cite{12}, etc., is the presence of second magnetization peak (SMP) anomaly in the isothermal scans at $T/T_c(0) < 0.6$. The field values of this SMP anomaly do not appear to vary with temperature. Prior to such a result in Ca$_3$Rh$_4$Sn$_{13}$, the SMP anomaly located deeper inside the mixed phase, which is distinct from the PE anomaly could be noted in the data of some of the crystals of high $T_c$ YBa$_2$Cu$_3$O$_7$ compound \cite{14,15,16,17}, however, the loci of SMP/PE fields in these samples have very different (complex and non-monotonic) behavior. Subsequent investigations in some of the weakly pinned crystals of $2H$-NbSe$_2$, LuNi$_2$B$_2$C, YNi$_2$B$_2$C, etc., have also revealed the presence of a tiny SMP anomaly distinct from the PE in isothermal magnetization hysteresis scans \cite{18,19,20}. The observations related to the SMP and the PE in the crystal of Ca$_3$Rh$_4$Sn$_{13}$ remain a spectacular exception. The locus of the SMP anomaly in it per se has close resemblance to that of the SMP anomaly in an `as grown' crystal of the compound Bi$_2$Sr$_2$CaCu$_2$O$_8$ (Bi2212), studied by Khaykovich {\it et al} \cite{21}. The SMP anomaly in Bi2212 is considered to be caused by the first order transition from an ordered Bragg glass (BG) state to the disordered vortex glass (VG) state. The PE anomaly in the crystals of low $T_c$ $2H$-NbSe$_2$ \cite{22}, Nb \cite{23}, etc., is also shown to fingerprint the first order like transition between the ordered and the disordered phases, such that one can easily encounter the superheating/supercooling effects across this transition as well as witness the presence of an interface separating weaker and stronger pinned regions above the onset temperature of the PE in the imaging experiments \cite{22}.

The notion of the phase coexistence across the first order transition(s) offers an interesting possibility of modulation in the phase coexistence characteristic across the SMP and the PE anomalies in isothermal scans in the context of our sample of Ca$_3$Rh$_4$Sn$_{13}$ \cite{8,9,11}. Park {\it et al} \cite{24} have recently drawn attention to the presence of multi-critical point in the BG phase in a single crystal of Nb, which had earlier been studied by Ling {\it et al} \cite{23}. Simultaneous measurement of small angle neutron scattering (SANS) and bulk ac susceptibility in a crystal of Nb have shown the observation of a BG phase below 0.8~kOe. However, this BG state does not yield a PE anomaly. The BG phase at $H < 0.8$~kOe reaches upto the superconducting phase boundary without encountering any order-disorder phase transition {\it a la} the PE. This facet of the data of Park {\it et al} \cite{24} could be termed as consistent with the earlier observation of Forgan {\it et al} \cite{25} at a somewhat higher field in a SANS study in another crystal of Nb, which the latter report asserted as being more pure than the Nb crystal used by Ling {\it et al} \cite{23}.

A motivation of this report on Ca$_3$Rh$_4$Sn$_{13}$ is to explore the evidence for modulation in a parameter which purports to measure the notion of phase coexistence across the SMP and the PE anomalies. Another objective is to present the results of an exploration to study the dynamical response in an isothermal $M-H$ scan via the dependence of the hysteresis width ($\Delta M(H) \propto J_c(H)$) on the ramp rate of the magnetic field ($\frac{dH}{dt}$). During the ramp up of the dc field in magnetization hysteresis experiments, the disordered bundles of vortices have to reach their (quasi) equilibrium configuration and thereafter, also attempt to participate in the order-disorder transition in an appropriate field region \cite{26,27}. The dynamical response at different field regions in the sample could carry integrated information on all underlying processes. Finally, we re-examine here the vortex phase diagram in the given Ca$_3$Rh$_4$Sn$_{13}$ sample in the context of the notion of the phase coexistence across the anomalous variation(s) in $J_c(H)$ and the possibility of (multi) critical points in the BG phase.

\section{Experimental details}

The Ca$_3$Rh$_4$Sn$_{13}$ crystal piece is the same ($T_c(0)\approx 8.2$~K) as studied by us earlier \cite{8}. While the ac susceptibility measurements were performed using a home built double coil set up, incorporating a mutual inductance bridge \cite{28}, the magnetization hysteresis ($M-H$) measurements were performed on a commercial vibrating sample magnetometer (Oxford Instruments, UK), in which the ramp rate of the magnetic field could be varied over a wide range. We had earlier obtained \cite{8} magnetization hysteresis data in this sample using a commercial SQUID magnetometer (Quantum Design Inc., U.S.A., Model MPMS5) and employing the procedure of the half scan, as enunciated by Ravikumar {\it et al} \cite{29,30}, to overcome a problem originating from the magnetic field inhomogeneity along the scan length.

\section{Results and Discussions}
\subsection{Second Magnetization Peak anomaly and the Peak Effect in the magnetization hysteresis loops}

Figure 1 shows portions of the $M-H$ plots obtained using a VSM at a few representative temperatures ({\it viz.}, 1.7~K, 4.3~K and 6.1~K), while sweeping the field at a rate of 2~kOe/min. The shapes of the $M-H$ loops displayed in figure~1 do not depend on the scan rate of the magnetic field. The $M-H$ data recorded on a SQUID magnetometer, where the magnetization values are recorded after putting the magnet current in the persistent mode, essentially reproduced the behavior recorded using a VSM (all data not being illustrated here). Note that the $M-H$ loop at 1.7 K (cf. figure~1(a)) displays the SMP anomaly as distinct from the PE. While the latter just precedes the approach to the irreversibility field (and the upper critical field, $H_{c2}$), the former anomaly is located deeper inside the mixed phase. In the given sample, the SMP anomaly commences somewhat below 10~kOe and its peak field is centered around 16~kOe. As the temperature is raised from 1.7~K to 4.3~K, the $H_{c2}$ decreases and the PE preceding it faithfully follows it. However, the field region of the SMP anomaly seems to be insensitive to the enhancement in temperature. At 4.3~K, the SMP anomaly can be seen to merge into the PE such that only one composite anomalous variation in the hysteresis width ($\propto J_c$) is evident in figure~1(b). In panel (a) of figure~1, if SMP is designated as the second peak (II), the PE anomaly would be the third peak (III) at 1.7~K. On warming up to 4.3~K, the composite anomaly has the status of the second peak (II). The $M-H$ data in the panel (c) of figure~1 displays the behavior at 6.1~K, which looks like the quintessential sharp PE bubble centered around 10~kOe. This peak can be termed as the second peak (II) in the $M-H$ loop at 6.1~K. At $T > 4.5$~K, only two peaks are evident in the $M-H$ loops of the given Ca$_3$Rh$_4$Sn$_{13}$ sample. In the initial phase of the investigations on this crystal, where the measurements were made above 4.2~K \cite{1,8}, the possibility of the presence of three peaks in $M-H$ data did not receive attention. The notion of two peaks remains evident upto a reduced temperature ($T/T_c(0)$) of about 0.94. Sarkar {\it et al} \cite{9} had shown that at $t = 0.95$ (7.83~K), the hysteresis width ($\propto J_c(H)$) collapses sharply near 2~kOe, without the display of an upturn at the edge of the collapse.

\subsection{Isothermal and Isofield in-phase ac susceptibility measurements}

Figure~2 provides a glimpse into the representative isothermal and isofield ac susceptibility data in Ca$_3$Rh$_4$Sn$_{13}$.  The panel (a) in figure~2 shows the field dependence of the in-phase ac susceptibility ($\chi^{\prime}(H)$) data at 4.3~K, where the field values corresponding to the onset of the SMP ($H_{smp}^{on}$) and the maximum of the PE ($H_{p}$) are marked  (cf. data in panel (a) of figure~1). As the temperature is increased to 6~K, $\chi^{\prime}(H)$ plot displays only a sharp peak centred around the $H_{p}$ value of 10~kOe (data not shown here). The panel (b) in figure~2 shows the isofield ac susceptibility ($\chi^{\prime}(T)$) data in a dc field of 10~kOe and measured with an imposed ac field of $h_{ac}=1$~Oe(r.m.s.) at driving frequencies of 21~Hz and 211~Hz, respectively. One can immediately observe the presence of two discontinuous jumps at temperatures marked as onset ($T_p^{on}$) and the peak temperature ($T_p$) of the PE phenomenon in the $\chi\prime(T)$ data. The ordered vortex lattice prepared in a field of 10~kOe (after zero field cooling) undergoes a stepwise disordering across the PE region \cite{8}. It has been pointed out earlier \cite{8} that the thermomagnetic history dependence in $\chi^{\prime}(T)$ response at $H=10$~kOe becomes indiscernible a little above $T_p$. The frequency dependence in $\chi^{\prime}(T)$ response becomes immeasurable (panel (b) of figure~2) at the limiting temperature $T^*$ ($>T_p$). In the notion of the phase co-existence across the temperature regime of the PE phenomenon, it is tempting to identify the interval between $T_p^{on}$ and $T^*$ as the region, where the superheated ordered domains dynamically coexist with the strongly pinned pockets. The temperature $T^*$ presumably marks the limit above which the sample is entirely filled with the disordered vortex matter and the dynamical effects related to the transformation between the ordered and the disordered regions cease. $T^*$ thus can be designated as the {\it spinodal temperature}, as asserted in the context of the analysis of the data in the weakly pinned crystals of 2$H$-NbSe$_2$ \cite{19,31} and Yb$_3$Rh$_4$Sn$_{13}$ \cite{32}. In 2$H$-NbSe$_2$, it has been shown that the $J_c(H)$ is path independent above $T^*$ \cite{19} and it rapidly decreases monotonically to the depinning limit at the irreversiblility temperature $T_{irr}$.

\subsection{Minor Hysteresis Curves: Thermomagnetic history dependence across the SMP and the PE}

When the SMP anomaly and the PE are present in an isothermal scan, the order-disorder transformation can be considered to commence at the onset field ($H_{smp}^{on}$) of the SMP anomaly. If the disordered stronger pinned pockets get nucleated and co-exist with the weaker pinned ordered regions above $H_{smp}^{on}$, then it is pertinent to ask as to how the disordered regions evolve in response to the increase in field above the peak field of the SMP anomaly ($H_{smp}$). Between $H_{smp}$ and the onset field of the PE ($H_p^{on}$), the width of the hysteresis loop ($\propto J_c(H)$) decreases as $H$ increases (see figure~1(a)), thereby implying that the interaction effects promote the spatial order in the vortex solid in this interval.

The $J_c(H)$ is known to imbibe thermomagnetic history dependence in response to the superheating/supercooling effects across the first order like order-disorder phase transition(s). The path dependence in $J_c(H)$ vividly manifests itself in isothermal magnetization hysteresis loops via the tracings of the minor hysteresis loops \cite{9,33}. When $J_c(H)$ is path independent and single valued function of $H$, the hysteresis loop obtained by cycling the magnetic field between $\pm H_{c2}$ defines an envelope contour within which lie all the minor hysteresis curves traceable in any chosen manner. However, when $J_c(H)$ becomes path dependent, the above premise does not hold \cite{33}. In Ca$_3$Rh$_4$Sn$_{13}$, the following inequality, $J_c^{FC}(H) > J_c^{rev}(H) > J_c^{for}(H)$, has been observed \cite{9}. The minor hysteresis  curves originating from $M_{FC}(H)$ cut across the envelope hysteresis loop, whereas those originating from $M^{for}(H)/M^{rev}(H$) undershoot/overshoot the reverse/forward leg of the envelope loop \cite{9,33}. The notion of undershooting connects to the superheating of the ordered state above the onset field of the order-disorder transformation during the ramp up of the field. 

 Kokkaliari's {\it et al} \cite{34,35} had introduced a procedure of examining and comparing saturated values of the minor hysteresis loops originating from $M^{for}(H$) values separated by equidistant field values ($\Delta H$). This procedure amounts to determining the difference ($\Delta M_{suc}$) in the magnetization values on the successive minor curves and plotting them as a function of $H$. The non-zero value of $\Delta M_{suc}$ implies a progressive enhancement in the plasticity ({\it i.e.} nucleation of disorder) of the ordered vortex solid. It is also instructive to gain information on the state of the spatial order of the vortex solid at a given $H$ on the envelope curve in comparison to the maximally disordered (metastable supercooled) state possible at the same $H$. Extending the reasoning of Kokkaliari's {\it et al} \cite{35}, the above can be accomplished \cite{36} if we examine the difference between the saturated value of the minor curve originating from a given $M_{FC}(H)$ value with the corresponding value of the envelope hysteresis curve and suitably normalize it to the width of the hysteresis loop, $\Delta M_{H}$. 

Figure 3(a) shows the field dependence of the differences ($\Delta M_{suc}$) in magnetization values (at a given $H$) on successive minor curves originating from $M^{for}(H)$ in Ca$_3$Rh$_4$Sn$_{13}$ at 1.7~K, where both SMP anomaly and PE are present. The data points in this plot are sparse, however, the trend is self evident. $\Delta M_{suc}$ {\it vs.} $H$ plot implies that the disordering of the flux line lattice ( {\it a la} Bragg glass phase) commencing at $H_{smp}^{on}$ reaches a limit near $H_{smp}$, and thereafter the disordering  again re-initiates at $H_p^{on}$ and it eventually ceases near the peak field $H_p$ of the PE. A key query as to whether (some of) the dislocations injected into the Bragg glass state between $H_{smp}^{on}$ and $H_{smp}$ are able to heal upto the onset field $H_p^{on}$, is addressed by the data presented in the panel(b) of figure~3.
 
Figure 3(b) depicts the field variation of the parameter $R_{FC}$ ($ = M_{FC}^{sat}(H)-M^{rev}(H)/ \Delta M(H)$), where $M_{FC}^{sat}$ is the saturated value of the magnetization on the FC minor curve and $M^{rev}(H)$ is the magnetization value on the reverse leg of the envelope loop at the corresponding $H$ value. As stated earlier \cite{36}, $R_{FC}$ parameterizes the difference between the notional ordered state and the notional maximally disordered state for a given $H$ ({\it i.e.}, for a given inter-vortex spacing $a_0$, as $a_0 \propto H^{-0.5})$. In figure~3(b), while reducing the field from highest field end, note that the parameter $R_{FC}$ starts to become finite as $H$ falls below the peak field $H_p$. In Ca$_3$Rh$_4$Sn$_{13}$, the disordering of the ordered lattice is nearly complete as $H \rightarrow H_p$ and above this field the disordered vortex solid is in equilibrium. The parameter $R_{FC}$ shows two maxima near $H_p^{on}$ and $H_{smp}^{on}$, thereby implying that the differences between the meta-stable supercooled state and the equilibrium ordered state maximize near the said field values. In between $H_p^{on}$ and $H_{smp}^{on}$, the parameter $R_{FC}$ shows a local minimum at the field value $H_{smp}$. This behaviour can be considered to imply that some of the dislocations interjected between $H_{smp}^{on}$ and $H_p^{on}$ indeed heal up, while ramping the field upto $H_p^{on}$. It may be emphasized that the vortex state obtained at $H=H_{smp}$ (while tracing the $M-H$ loop at 1.7~K) is only partially disordered and it is distinctly different from the meta-stable disordered state prepared in the FC manner. The state of co-existence between $H_{smp}^{on}$ and $H_p^{on}$ shows a modulation, however the ordered regions appear to dominate the composite response of the sample. Above $H_p^{on}$ , the proliferation of dislocations into the sample rapidly transforms it towards a state which is disordered in equilibrium and there is no further possibility of any healing of the dislocations injected into the sample. 

 In figure~3(b), it can also be noted that the parameter $R_{FC}$ assumes negative values as $H$ reduces below 8~kOe. This implies that the vortex state obtained while increasing the field ({\it i.e.} the ZFC state) in an isothermal manner is less ordered than that obtained while field cooling in $H$ values lying between 3~kOe and 6~kOe. Such an observation can be rationalized by recalling \cite{26} that while ramping the field, the vortex bundles presumably get injected in the sample through broken edges, corners, imperfections, etc. in a disordered manner and they dynamically heal into an ordered state in response to interaction effects between the vortices. At low field values, the interaction effects are not adequate to completely overcome the disorderliness induced by sample edges (edge contamination \cite{26}). The vortex state prepared in the field cool manner is not susceptible to menace of edge contamination and, therefore, at low fields, we witness the inequality, $J_c^{FC}(H) < J_c^{rev}(H) < J_c^{for}(H)$ (all data not shown here, see figure~3 in Ref [36]. 

\subsection{Ramp rate dependence of the hysteretic magnetization response: Modulation in dynamical response across the SMP and the PE}

As stated in the introduction (section 1), it could be instructive to examine the ramp rate dependence of $J_c(H)$ (proportional to hysteresis width of the $M-H$ loop at a given field $H$) to gain information on the time scale at which disorder injected through the edges dynamically anneals towards the equilibrium vortex state at a given $H$. We had recorded $M-H$ loops at  varying ramp rates (6~kOe/min. to 0.1~kOe/min.) at a few selected temperatures. Figures 4(a) and 5(a) show the portions of the half-widths of hysteresis loop pertaining to the forward leg of the envelope curve. Such half-width values correspond to critical current density in the sample durng the so called zero field cooled state, which we designate as $J_c^{for}(H)$. From the curves in figures 4(a) and 5(a), it is apparent that the ramp rate dependence of $J_c^{for}$ values is different in different field regions of the hysteresis loop. $J_c^{for}(H)$ can be seen to display larger dependence on the ramp rate at the low field end ($H < H_{smp}^{on}$), and it decreases as $H$ exceeds $H_p^{on}$. In between $H_{smp}^{on}$ and $H_p^{on}$, the dependence of $J_c(H)$ on the ramp rate shows an interesting modulation, as detailed below.

 We ascertained that the normalizated $J_c^{for}(H)$ values at a given $H$ have a logarithmic dependence on the inverse of the ramp rate. The $J_c^{for}$ values have been normalized to an extrapolated value for the infinite ramp rate ($J_c^\infty$) for our purpose here. From such plots, we could compute the magnitude of the slope parameter $S$, which is proportional to $-(J_c^{\infty})^{-1}\frac{d(J_c^{for}(H))}{d(log(ramp~rate^{-1}))}$. The data points in figures 4(b) and 5(b) show the variation of $S$ {\it vs.} $H$ at 1.9~K and 3.86~K, respectively. Before we proceed to discuss this behaviour, it may be stated that the local field in the sample could vary from the edges of the sample to its interior, as per standard prescription of the Bean$^{\prime}$s Critical State Model \cite{37}. However, the values of the gradient in field across the sample at different $H$ can be estimated by tracings of the minor hysteresis loops \cite{9} initiated from different $H$. We reckon that these values are of the order of 10$^2$~Oe in the field region of figures 4(b) and 5(b). Therefore, to a first approximation, we may assume that the average field across the sample is not very different from the applied field $H$.

 The qualitative trend in figures 4(b) and 5(b) is that the $S$ values are large prior to $H_{smp}^{on}$, and they show an upward trend upto $H_{smp}^{on}$. The turnaround in $S$ values above $H_{smp}^{on}$ is followed by a gradual decrease to a (local) minimum value near the peak field $H_{smp}$ of the SMP anomaly. Above $H_{smp}$, the $S$ parameter enhances again to reach a local maximum near $H_p^{on}$, and thereafter, it decreases again as $H \rightarrow H_p$. The nature of the modulation in $S$ bears close resemblance to the variation in $R_{FC}$ parameter displayed in figure~3(b). In view of this similarity, it is tempting to visualize that the extent of spatial order in the equilibrium vortex state at a given $H$ dictates the ramp rate dependence of the hysteretic magnetization response. When the destined  vortex state is well ordered in equilibrium, the injected disorder requires a larger time to anneal towards it. The instantaneous magnetization value at a given $H$ during high ramp rate is larger, thereby, reflecting the incomplete annealing of the stronger pinned disordered packets towards the ordered matrix. Another extreme is when the destined vortex state is disordered in equilibrium. In such a situation, the approach to the destined state happens faster, and the $S$ parameter has a minimum value. This minimum value presumably corresponds to the usual long time temporal decay of $J_c$ due to the notion of flux creep in pinned superconductors. In between the two extreme circumstances, the modulation in the values of the $S$ parameter could be interpreted to measure the partial order in the state of co-existence of the ordered and the disordered regions. The observed increase in the value of the $S$ parameter between $H_{smp}$ and $H_p^{on}$ reinforces the assertion made in section 3.3 that the state of spatial order improves in this field interval. 

\subsection{Summary: Critical points in the updated vortex phase diagram of Ca$_3$Rh$_4$Sn$_{13}$}

 To summarize, it is useful to reconstruct the vortex phase diagram in Ca$_3$Rh$_4$Sn$_{13}$, keeping in view the notion of phase coexistence above the onset position of order-disorder transition(s). In isothermal scans at $T < 5$~K, one can locate two field positions ($H_{smp}^{on}$ and $H_p^{on}$) at which the order-disorder transitions commence. As the temperature exceeds 5~K, the SMP anomaly juxtaposes with the PE and one can mark out only a composite onset position of the order-disorder transition; such a position has a continuity with the onset field values ($\sim 9$~kOe) of the SMP anomaly, which show little variation with temperature from 1.7~K to 5~K. Figure 1(c) shows that at 6.1~K, one can witness only the PE anomaly, whose onset position is not very different from $H_{smp}^{on}$ value of $\sim 9$~kOe. Above 6.1~K, the onset field of the PE decreases as the temperature increases.

 We show in figure~6 an updated version of the vortex phase diagram in the given crystal of Ca$_3$Rh$_4$Sn$_{13}$, which incorporates the above description. This phase diagram comprises the plots of $H_{smp}^{on}$, $H_p^{on}$, $H_{irr}$, $H_{sb}$ and $H_{c2}$. The $H_{sb}$ line identifies the limit of the small bundle pinning regime, as described earlier (see figure~6 in Ref. [9]). Above this line, the ordered vortices are expected to be collectively pinned in large domains. The region of large domains gradually transforms into the elastic glass ({\it i.e.}, Bragg glass state) phase in a continuous manner. The $H_{irr}$ line passes through the limiting field/temperature values at which pinning in the bulk of the crystal ceases. The spinodal line ($H^*$ line), which could identify the limiting fields/temperatures at which the phase coexistence ceases, lies a little inside the $H_{irr}$/$T_{irr}$ line. (cf. figure~2(b)). Hence for notional purpose, we may take the irreversibility line to identify the spinodal ($H^*,T^*$) line as well in the present crystal of Ca$_3$Rh$_4$Sn$_{13}$.

 In figure~6, the phase coexistence $H,T$ region between the $H_{smp}^{on}$ line and the $H_{irr}$ line has been shown to bifurcate into two parts (I and II), in which the ordered and the disordered regions dominate the bulk magnetic response, respectively. Such a sub-division could be termed as consistent with an analysis presented for the vortex phase diagram in a weakly pinned crystal of 2$H$-NbSe$_2$, which also yielded SMP anomaly as distinct from the PE \cite{19,31}. Note that in Ca$_3$Rh$_4$Sn$_{13}$, the $H_{smp}^{on}$ line and the $H_p^{on}$ line have been shown to meet at the multi-critical point A and the composite line at higher temperatures ({\it i.e.}, the $H_p^{on}$ line) ends in a critical point B (see figure~6). The ($H,T$) region between the $H_{sb}$ line and the $H_{smp}^{on}$ line is also identified as the elastic glass (Bragg glass) state. The ordered Bragg glass phase yields a PE feature in temperature variation of the ac susceptibility at $H \ge 3.5$~kOe \cite{8}. However, in the ($H,T$) region below the critical point B, it has not been possible to observe the PE feature even in isothermal $M-H$ scans. We recall here that the hysteresis width at 7.83~K (see figure~8 in Ref. [9]) displays a sharp collapse near 2~kOe, without the PE feature. 

 In conclusion, we may redraw attention to the notion of multi-critical point in the Bragg glass state as introduced by Park et al \cite{24}, while discussing their data in a weakly pinned crystal of Nb. The multi-critical point A in Ca$_3$Rh$_4$Sn$_{13}$ has a different physical basis as compared to that in the crystal of Nb. The critical point B in Ca$_3$Rh$_4$Sn$_{13}$ could be a  counterpart of the multi-critical point referred to by Park et al \cite{24} in Nb. In a crystal of Ca$_3$Rh$_4$Sn$_{13}$, there is no evidence of the phenomenon of surface superconductivity, which yielded the notion of multi-critical point in Nb.

\section{Acknowledgment}

Two of us (SS and ADT) would like to acknowledge the TIFR endowment fund for the Kanwal Rekhi career development support.

$^\dagger$ Email: grover@tifr.res.in~~:~~$^{\ast}$ Email: tomy@phy.iitb.ac.in

\newpage

\begin{figure} %[h]   % Figure~1  
\includegraphics[scale=1.0,angle=0]{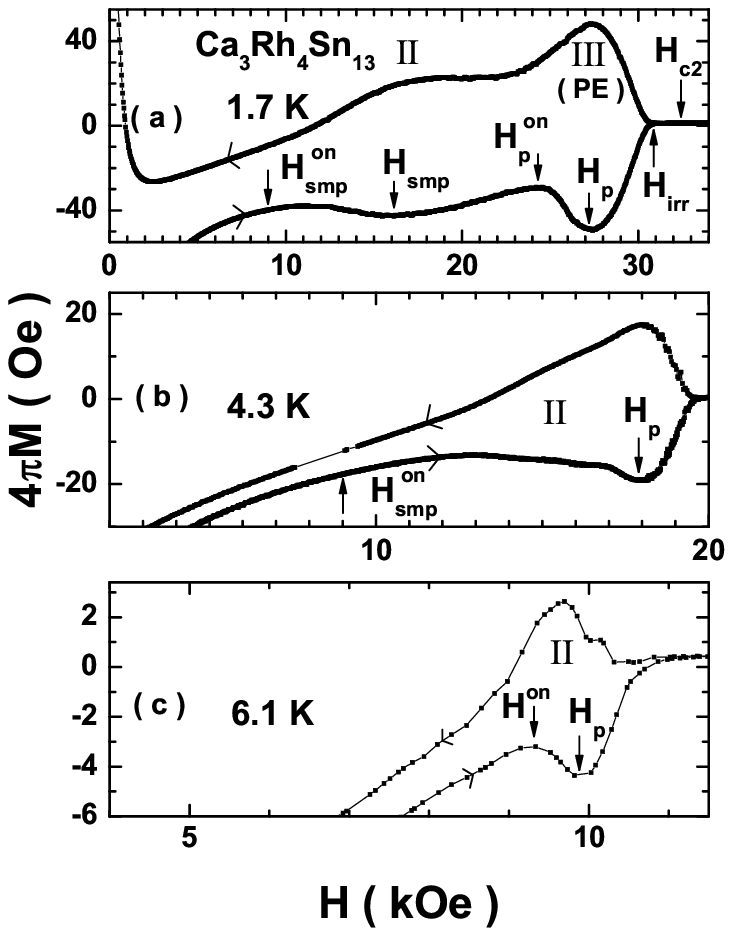}
\caption{ Portions of the isothermal magnetization hysteresis loops showing second magnetization peak (SMP) anomaly and/or the peak effect (PE) in a single crystal of Ca$_3$Rh$_4$Sn$_13$ at (a) 1.7~K, (b) 4.3~K and (c) 6.1~K}
\end{figure}
\begin{figure} %[h]   % Figure~2  
\includegraphics[scale=1.0,angle=0]{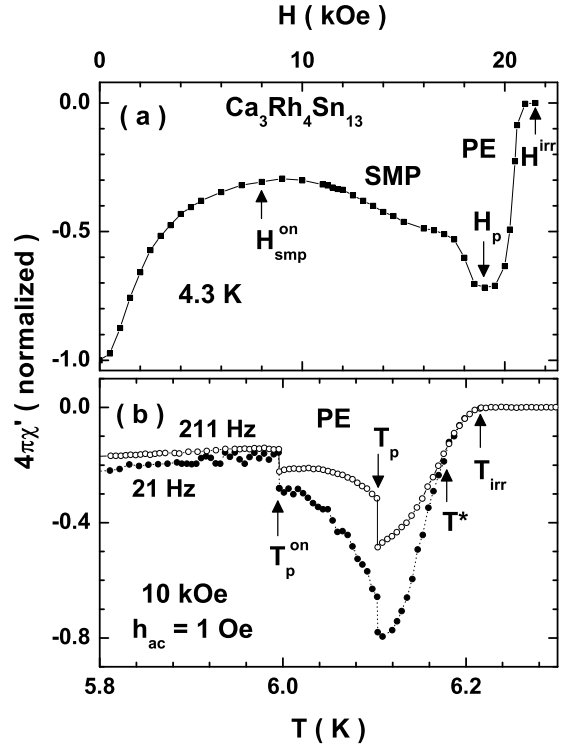}
\caption{ Examples of isothermal (panel (a)) and isofield (panel (b)) in-phase ac susceptibility ($\chi\prime_{ac}$) data in a crystal of Ca$_3$Rh$_4$Sn$_{13}$. The positions of onset field of the SMP anomaly, $H_{smp}^{on}$, and that of the peak field of the PE, $H_p$, at 4.3~K have been identified in panel (a). The isofield data at $H=10$~kOe in panel (b) shows the notion of step-wise amorphization triggered at $T_p^{on}$ and $T_p$ values. The limiting temperature ($T^*$) above which the $\chi\prime_{ac}$ responses at 21~Hz and 211~Hz become nearly identical has been marked.   }
\end{figure}
\begin{figure} %[h]   % Figure~3  
\includegraphics[scale=1.0,angle=0]{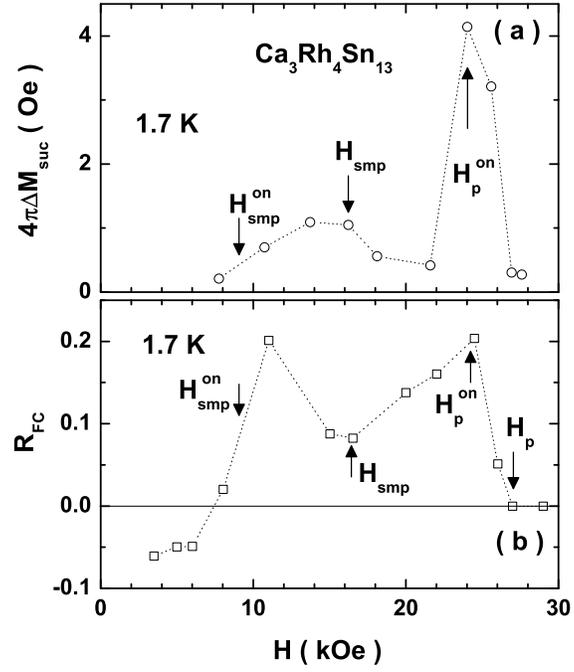}
\caption{ The panel (a) shows the plot of difference magnetization, $\Delta M_{suc}$, {\it vs.} field at 1.7~K in a crystal of  Ca$_3$Rh$_4$Sn$_{13}$. The values of  $\Delta M_{suc}$  have been obtained from the tracings of the minor hysteresis loops originating from selected magnetic fields differing by a chosen interval. The panel (b) depicts the plot of the parameter $R_{FC}$ {\it vs.} field (see text for detail) at 1.7~K.}
\end{figure}
\begin{figure} %[h]   % Figure~4  
\includegraphics[scale=1.0,angle=0]{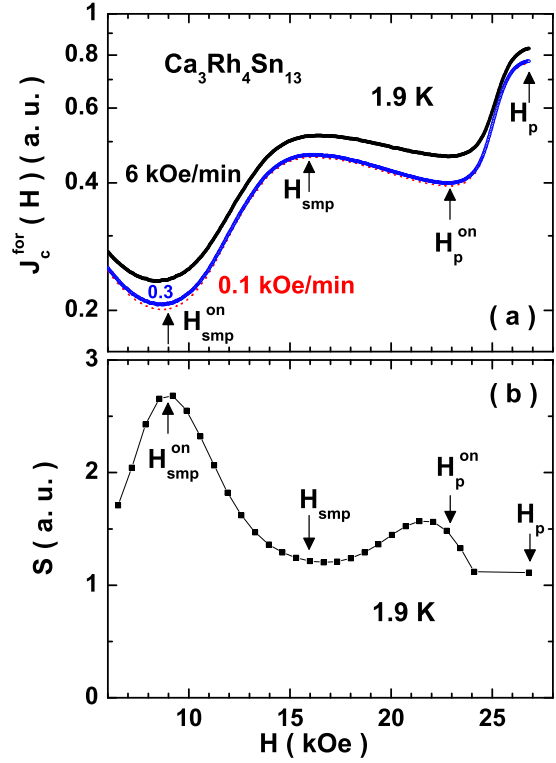}
\caption{The panel (a) depicts the plots of the field dependences of the critical current density $J_c^{for}(H)$, corresponding to the forward leg of the magnetization hysteresis loops, recorded at different scan rates in the crystal of Ca$_3$Rh$_4$Sn$_{13}$ at 1.7~K. Different curves are labelled by the respective values of the scan rate. The panel (b) shows the field dependence of the slope parameter $S$ at 1.7~K. The $S$ parameter value at each field has been extracted from an analysis of the ramp rate dependence of the $J_c^{for}(H)$ data (see text for detail). The locations of the onset and peak field of the SMP anomaly and those of the peak effect at 1.7~K have been identified in both the panels.
}
\end{figure}
\begin{figure} %[h]   % Figure~5  
\includegraphics[scale=1.0,angle=0]{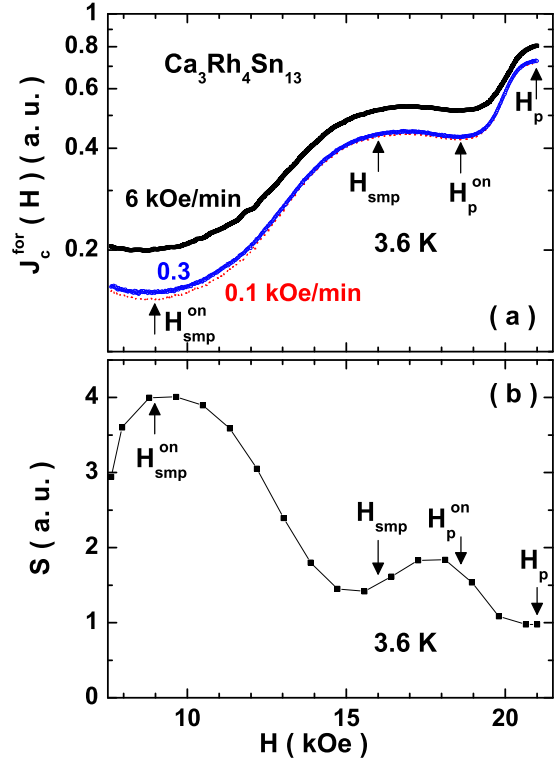}
\caption{The panel (a) shows the plots of the field dependences of $J_c^{for}(H)$ for different scan rates in a crystal of Ca$_3$Rh$_4$Sn$_{13}$ at 3.6~K. The panel (b) shows the behaviour of the slope parameter $S$ versus field at 3.6~K. The positions of the onset and the peak field of the SMP anomaly and those of the PE at 3.6~K have been marked in both the panels.}
\end{figure}
\begin{figure} %[h]   % Figure~6  
\includegraphics[scale=0.5,angle=0]{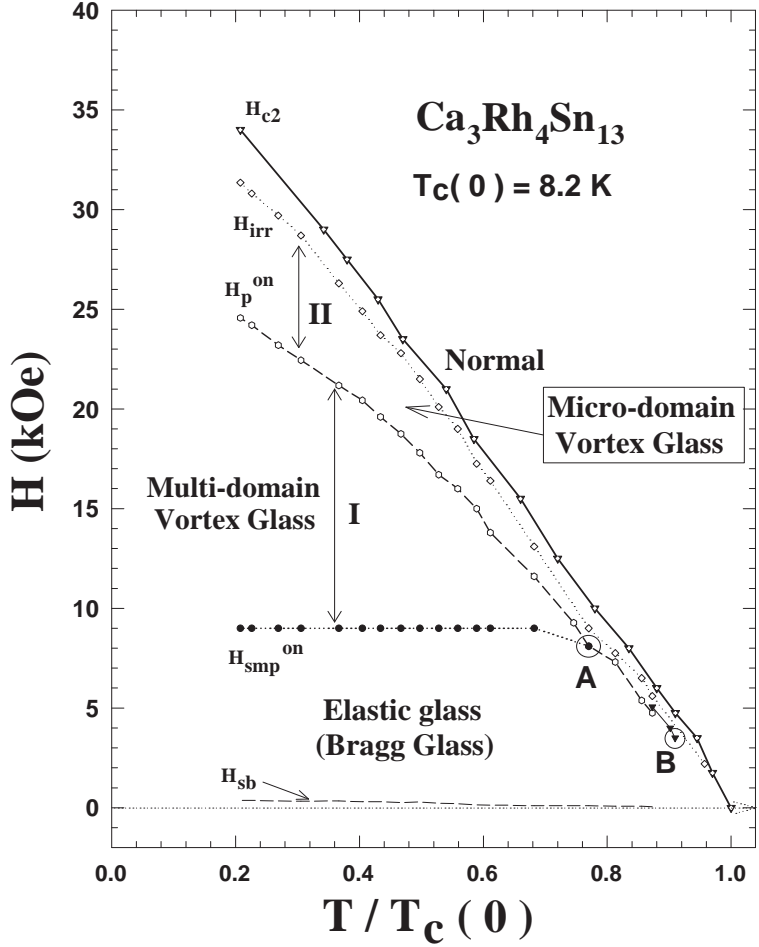}
\caption{Vortex phase diagram for the weakly pinned crystal of Ca$_3$Rh$_4$Sn$_{13}$. We have plotted the data points corresponding to the onset fields of the SMP anomaly and the PE along with the irreversibility field from the isothermal $M-H$ hysteresis loops. The values of the (limiting) small bundle pinning fields ($H_{sb}$) were obtained from an analysis of $J_c(H)$ {\it vs.} $H$ data as in Ref. [9]. The vortex phases in different ($H,T$) regions have been identified. The points A and B identify the multi-critical point and the critical point, respectively.}
\end{figure}
\end{document}